\documentclass[twocolumn,showpacs,preprintnumbers,amsmath,amssymb]{revtex4}
\newcommand{\be}{\begin{equation}}
\newcommand{\ee}{\end{equation}}
\newcommand{\bea}{\begin{eqnarray}}
\newcommand{\eea}{\end{eqnarray}}
\newcommand{\ba}{\begin{array}}
\newcommand{\ea}{\end{array}}

\def\bbox{{\,\lower0.9pt\vbox{\hrule \hbox{\vrule height 0.2 cm
\hskip 0.2 cm \vrule height 0.2 cm}\hrule}\,}}
\newcommand{\dsl}{\pa \kern-0.5em /}

\newcommand{\nn}{\nonumber \\}

\def\0{{\sst{(0)}}}
\def\1{{\sst{(1)}}}
\def\2{{\sst{(2)}}}
\def\3{{\sst{(3)}}}
\def\4{{\sst{(4)}}}
\def\5{{\sst{(5)}}}
\def\6{{\sst{(6)}}}
\def\7{{\sst{(7)}}}
\def\8{{\sst{(8)}}}

\def\ep{\epsilon}

\let\a=\alpha \let\b=\beta   
  \let\q=\theta  
 \let\m=\mu   
   \let\f=\phi

\let\la=\label  
  
\def\nn{\nonumber} \def\bd{\begin{document}} \def\ed{\end{document}}
\def\ds{\documentstyle} \let\fr=\frac \let\bl=\bigl \let\br=\bigr
\let\Br=\Bigr \let\Bl=\Bigl
\let\bm=\bibitem
\let\na=\nabla
\let\pa=\partial \let\ov=\overline
\def\ft#1#2{{\textstyle{{\scriptstyle #1}\over {\scriptstyle #2}}}}
\def\fft#1#2{{#1 \over #2}}
\def\del{\partial}
\def\sst#1{{\scriptscriptstyle #1}}
 \def\oneone{\rlap 1\mkern4mu{\rm l}}
\def\ie{{\it i.e.\ }}
\def\via{{\it via}}
\def\semi{{\ltimes}}
\def\str{{\rm str}}
\def\Dm{{{D_{\sst{max}}}}}
\def\vac{ \left | 0 \right \rangle }
\def\kvac{ \left | k \right \rangle }

\newcommand{\hsp}{\hspace{0.5cm}}

\usepackage{graphicx}% Include figure files
\usepackage{dcolumn}% Align table columns on decimal point
\usepackage{bm}% bold math

\begin{document}

\preprint{ITFA-2006-33, hep-th/0609154}

\title{Fuzzball solutions for black holes 
and D1-brane-D5-brane microstates}%Force line breaks with \\

\author{Kostas Skenderis${}^*$ and Marika Taylor\footnote{The authors
are supported by NWO.}}
% \altaffiliation[Also at ]{Physics Department, XYZ University.}%Lines break automatically or can be forced with \\
%\author{Second Author}%
% \email{Second.Author@institution.edu}
\affiliation{%
Institute for Theoretical Physics, University of Amsterdam,\\
Valckenierstraat 65, 1018 XE Amsterdam, The Netherlands.
%This line break forced with \textbackslash\textbackslash
}%

%\date{\today}% It is always \today, today,
             %  but any date may be explicitly specified

\begin{abstract}
We revisit the relation between fuzzball solutions and D1-D5
microstates. A consequence of the fact that the R ground
states (in the usual basis) are eigenstates of the R-charge is that
only neutral operators can have non-vanishing expectation values on
these states. We compute the holographic 1-point functions of the
fuzzball solutions and find that charged chiral primaries have
non-zero expectation values, except when the curve characterizing the
solution is circular. The non-zero vevs reflect the fact that a
generic curve breaks R-symmetry completely. This implies that
fuzzball solutions (excepting circular ones) can only correspond to
superpositions of R states and we give a proposal for the
superposition corresponding to a given curve. We also address the
question of what would be the geometric dual of a given R ground 
state.

\end{abstract}

\pacs{11.25.Tq\ \ 04.70.-s\ \ 11.25.-w}% PACS, the Physics and Astronomy
                             % Classification Scheme.
%\keywords{Suggested keywords}%Use showkeys class option if keyword
                              %display desired
\maketitle

Black holes provide a unique probe of non-perturbative aspects of
quantum gravity and as such they have played a prominent role in
many important developments over the years. In particular, the fact
that they carry entropy and Hawking radiate, along with the
associated information loss paradox, imposes restrictions and
challenges on any potential quantum theory of gravity. Progress in
resolving these issues has been achieved for a class of
supersymmetric black holes using string theory. The degeneracy of
the microstates of these black holes can be computed from first
principles using D-branes. Furthermore, the AdS/CFT correspondence
implies that the gravitational evolution is unitary. These
developments however have left unanswered the question of what are
the black hole microstates in the original gravitational description
of the system.

An interesting proposal for the gravitational nature of the
microstates was made in \cite{Lunin:2001jy}, see
\cite{Mathur:2005zp} for a review and further references. According
to this proposal there should be an exponential number of
horizon-free solutions associated to each black hole, one for each
microstate (although in general only a subset of these solutions
will be realized as good supergravity solutions). The microstates
should resemble the original black solution asymptotically and
(generically) they should differ from each other up to the horizon
scale. The original black hole would provide only the ``average''
description of the system; the true description is in terms of the
non-singular microstate geometries.

This is a very appealing picture that would resolve long standing puzzles 
associated with black hole physics such as the information loss paradox.
It is thus important to scrutinize the evidence for it and further 
develop it in a way that quantitative analysis is possible. 
Such a precise quantitative framework is required for this picture to 
become a physical model that would ultimately resolve the issues
discussed above or be falsified. 
The main aim of this work is to provide such a framework.

A test case for this proposal has been the 2-charge D1-D5 system.
This is a 1/4 supersymmetric system and the ``naive'' black hole
geometry has a near-horizon geometry of the form $AdS_3 \times S^3
\times M$, where $M$ is either $T^4$ or $K3$. 
%The naive geometry has
%a naked singularity but one expects that a horizon would emerge
%from $\a'$ corrections. 
Associated with this
system an exponential number of horizon-free solutions has been
found \cite{Lunin:2001jy} and proposed to correspond to microstates.
These solutions can account for a finite fraction of the black hole
entropy. One still needs an exponentially large number of solutions
to account for all of the black hole entropy and such solutions
were discussed in \cite{Lunin:2002iz}. 
%these solutions
%should depend on the structure of the internal manifold. Such
%solutions, related to the odd cohomology of $T^4$ were constructed
%in \cite{Taylor:2005db} and the construction of solutions associated
%with the middle cohomology of the internal manifolds was discussed
%in \cite{Lunin:2002iz}.
In this case the AdS/CFT correspondence provides
a dual description of the microstates so one should be able to
quantitatively understand the relation between the 
horizon-free geometries and the actual microstates.

Ultimately the fuzzball picture should apply to any black hole, 
supersymmetric or not. It would however be an enormous task to find
the required number of horizon-free solutions for all cases.
Even in the next simplest case, i.e. the 1/8 supersymmetric black 
holes obtained by adding momentum charge to the D1-D5 system,
the current results are very incomplete with only a small number 
of horizon-free geometries known. 
%Furthermore, the known non-supersymmetric ``fuzzball'' solutions 
%are classically unstable. 
From our point of view,
one should first understand in detail the precise properties
that the required solutions should have and how many 
of them should be 
accessible in supergravity 
%(as opposed to full fledged string theory)
before embarking upon such a task.
Thus one should first understand completely the D1-D5 system, 
in particular, how black hole properties emerge upon coarse graining.
 
The solutions of \cite{Lunin:2001jy} were found using the fact that
the D1-D5 system is related by dualities to a fundamental string
carrying momentum. Such F-P solutions were constructed in
\cite{Dabholkar:1995nc} and carrying out the dualities leads to the
following D1-D5 solution
 \bea ds^2 &=& f_{1}^{- 1/2} f_{5}^{- 1/2}
\left ( - (dt - A)^2 + (dy +
B)^2 \right ) \nn \\
&&+ f_{1}^{1/2} f_{5}^{1/2} dx \cdot dx +
f_{1}^{1/2} f_{5}^{- 1/2} dz \cdot dz; \nn \\
e^{2 \Phi} &=& f_1 f_5^{-1}; \hsp
C_{ti} = f_{1}^{-1} B_{i}; \hsp
C_{ty} = f_{1}^{-1}; \la{ez1} \\
C_{yi} &=& f_{1}^{-1} A_{i}; \hsp
C_{ij} = c_{ij} - f_{1}^{-1} (A_i B_j - A_j B_i), \nn
\eea
where the metric is in the string frame and the
fields satisfy
\bea
&&dc = \ast_{4} df_5, \hsp d B = \ast_{4} dA, \nonumber \\
&& \Box_{4} f_1 = \Box_{4} f_5 = \Box_{4} A_i =0, \qquad
\partial^i A_i = 0. \label{cond}
\eea
where the Hodge dual $\ast_4$ and $\Box_{4}$ are defined on the four
(flat) non-compact overall transverse directions $x^i$. The compact part
of the geometry can be either $T^4$ or $K3$.

The harmonic functions appearing in the solutions of \cite{Lunin:2001jy}
are determined by an arbitrary closed curve $F^i(v)$ of length $L$ in $R^4$,
\bea \la{mathu1}
&&f_{5} = 1 + \frac{Q_5}{L} \int^{L}_{0} \frac{dv}{\left | x - F \right
  |^2}; \
f_1 = 1 + \frac{Q_5}{L}
\int_{0}^L \frac{dv | \dot{F} |^2}{\left | x - F \right |^2}; \nn \\
&&A_i = \frac{Q_5}{L} \int_{0}^{L} \frac{\dot{F}_i dv}{\left | x - F
  \right |^2},
\eea
The physical interpretation of these solutions is that the D1 and D5
brane sources are distributed on a curve in the transverse $R^4$. The
D5-branes are uniformly distributed along this curve, but
the D1-brane density at any point on the curve depends on the tangent
to the curve, % The total one brane charge is given by
%\be \la{one-brane}
$Q_1 = (Q_5/L) \int_{0}^L | \dot{F} |^2 dv$,
%\ee
%Both the $Q_i$ have dimensions of length squared and 
where the $Q_i$ are related to
the integral charges by $Q_1 = (\a')^3 n_1/V, Q_5 = \a' n_5$,
with $(2 \pi)^4 V$ the volume of the compact manifold
(we set the string coupling constant to one, 
%string coupling constant, which plays no role in our discussion, to one,
$g_s=1$). Furthermore, the length of the curve is given by $L = 2 \pi Q_5/R,$
where $R$ is the radius of the $y$ circle. Finally, we require that
the curve satisfies
%\be \la{c2}
$\int dv F^{i} (1 + (Q_5/Q_1) | \dot F |^2 ) = 0$.
%\ee
This corresponds to choosing the origin of the coordinate system
to be at the center of mass of the system. 

It was argued in \cite{Lunin:2001jy} that these solutions
are related to the R ground states and via spectral flow also
to chiral primaries common to both the $T^4$ and $K3$ boundary CFTs
\cite{Lunin:2002bj}. Roughly speaking, the proposal was that
different curves correspond to different R ground states/chiral primaries.
We will show in this letter that only the circular curves can
correspond to R ground states/chiral primaries (in the usual basis)
and we will propose a precise map between all geometries and
superposition of R ground states. We further address the question 
of what would be 
the geometric dual of a given R ground state.

The low energy dynamics of the D1-D5 system
is governed by a 2d CFT  with ${\cal N}=(4,4)$ supersymmetry.
The internal R-symmetry is $SU(2)_L \times SU(2)_R$,
with $SU(2)_L$ carried by left movers and $SU(2)_R$ by right movers.
The detailed structure of the CFT will not be important in our discussion;
we will only need some general facts about the chiral primaries
and the associated R ground states that we now review.
The chiral primaries in the NS sector are states with
$h^{NS}=j^{NS}_3, \bar{h}^{NS}=\bar{j}^{NS}_3$,
where $(h^{NS},\bar{h}^{NS})$ are the conformal
dimensions  and $j_3^{NS}$ and $\bar{j}^{NS}_3$ are the
$U(1) \times U(1)$ charges (of the Cartan of $SU(2)_L \times SU(2)_R$)
for the left and right moving sectors, respectively.
For each chiral primary there is a corresponding R ground state
obtained via spectral flow, with
$h^R =h^{NS} - j_3^{NS} + c/24, \quad j_3^R=j_3^{NS} -c/12.$
The R-ground states thus obtained have the same dimension and
{\it definite} R-charges.

Consider now the expectation value of an operator
$O^{k_1,k_2}$ of charge $j_3^R=k_1, \bar{j}^{R}_3=k_2$
in a R ground state. Charge conservation
implies that this expectation value is zero, unless the operator
is neutral under $U(1) \times U(1)$,
\be \label{zerov}
\langle -j_3^R,-\bar{j}_3^R| O^{k_1,k_2} | j_3^R,\bar{j}_3^R \rangle =0,
\quad \{k_{1} \neq 0\ or\
k_{2} \neq 0\}
\ee
where we label the R ground states by their R charge and use the
fact that the bra state has the opposite R charge to the ket state.
It follows that for a geometry to be dual to a specific R ground state
the holographic 1-point functions of all charged operators must be
equal to zero.

In the decoupling limit, one may drop the one in the harmonic
functions (\ref{mathu1}) and the fuzzball solutions become asymptotically
$AdS_3 \times S^3$.  One can then use the AdS/CFT correspondence
to compute the expectation
value of 1/2 BPS operators of the dual CFT. A detailed analysis of this
computation appears in a companion paper \cite{KST}; here we
will only present the results for a few relevant operators.

Recall that in the AdS/CFT correspondence the boundary
fields parametrizing the Dirichlet boundary conditions of supergravity fields
are identified with sources of boundary gauge invariant operators
\cite{Gubser:1998bc}.
Thus to obtain the vev of a given operator
we need to compute the variation of the on-shell action w.r.t. the
corresponding source. This computation requires appropriate renormalization
and results in a concrete formula for the vev in terms of certain coefficients
in the asymptotic expansion of the bulk fields \cite{deHaro:2000xn}
(see \cite{Skenderis:2002wp} for a review and further references).
A precise holographic map between the vevs of 1/2 BPS operators
and asymptotically $AdS \times S$ solutions was constructed in our
recent work \cite{Skenderis:2006uy}. This construction
starts by decomposing the deviation of the fuzzball
metrics from $AdS_3 \times S^3$ in $S^3$ harmonics and performing
a radial expansion of the resulting coefficients.

The six dimensional theory obtained by reducing the ten dimensional solutions
over the internal manifold involves the metric $g_{MN}$, the dilaton
$\Phi$ and the 3-form (which we decompose into self-dual and
anti-delf dual parts) $G^{(\pm)}=1/2 (1\pm *)d C$.
Let us denote the deviations of the
six dimensional solution from the $AdS_3 \times S^3$ solution
by ($h_{MN}, g^{(\pm)}_{MNP}, \f$), where $M=\{\m,a\}$ and $\m$ and $a$
are $AdS_3$ and $S^3$ indices, respectively.  The harmonic expansion of these
deviations contains many terms; here we only give the terms
relevant for us
\bea
h_{\m a} &=& \sum h_{\m}^{I_v \pm} (x) Y_a^{I_v \pm} (y) + \cdots;  \nn \\
h_{(a b)} &=& \sum
\rho^{I} (x) D_{(a} D_{b)} Y^{I} (y)  + \cdots; \nn \\
h^{a}_{a} &=& \sum \pi^{I} (x) Y^{I} (y); \label{KK} \\
g^{(\pm)}_{\m a b} &=& \sum (D_{\m} U^{(\pm) I} \ep_{abc} D^{c} Y^I 
+
2 Z_{\m}^{(\pm) I_v \pm} D_{[b} Y^{I_v \pm}_{a]}); \nn \\
\phi  &=& \sum \phi^{I} (x) Y^{I} (y), \nn
\eea
where $x$ and $y$ are $AdS_3$ and $S^3$ coordinates, parenthenses denote
symmetrization with the trace removed, and  $Y^I$ and
$Y_a^{I_v \pm}$ are scalar and vector harmonics of $S^3$,
with the degree $k$ vector harmonics satisfying (apart from the
usual eigenvalue equation)
$\ep_{abc} D^b Y^{c I_v \pm} = \pm (k+1) Y_{a}^{I_v \pm}$.

The spectrum of the KK excitations was computed in \cite{Deger:1998nm}
and was matched to the spectrum of 1/2 BPS operators of the dual CFT in
\cite{deBoer:1998ip}. In particular, the KK spectrum consists of a tower of spin two supermultiplets,
along with two towers of spin 1 supermultiplets, one coming from
combinations of 
the scalar and the anti-self dual
tensor field and the second coming from the metric and self-dual tensor field.
The lowest lying chiral primaries of these two towers are
$S^1_j$, $\Sigma^2_I$ and they correspond respectively
to 4 scalar operators $O_{S^1_j}$
($j=1,\ldots,4$) of dimension 1 transforming in the $(1/2, 1/2)$ representation
of the R-symmetry $SO(4)$, and 9 scalar operators $O_{\Sigma^2_I}$ of dimension
2 transforming in the $(1,1)$ representation.
We will also be interested
in the R-symmetry current $j^{\pm \a}_u$, where $\a$ is an $SU(2)$
adjoint index
and $u$ a boundary spacetime index. This corresponds to the gauge
field $A^{\pm \a}_u$ coming from the spin two tower.

The vacuum expectation values of these operators
can be obtained from the asymptotics of the coefficients in (\ref{KK}).
%using the following holographic 1-point functions,
%\bea \label{1pt}
%\langle O_{S^1_j} \rangle &=& \frac{\sqrt{2} N}{4 \pi}
%[\f_j + 6 U^{(-)}_j]_{1}  \\
%\langle O_{\Sigma^2_I} \rangle &=& \frac{\sqrt{2} N}{\pi}
%\left[\frac{4}{3} U^{(+)}_I - \frac{1}{18} \pi_I - \frac{10}{9} \r_I
%\right. \nn \\
%&-&\left.
%\frac{1}{192} (\f_i + 6 U^{(-)}_i) (\f_j + 6 U^{(-)}_j) a_{Iij}
%\right]_{2} \nn \\
%\langle j_u^{\pm \a} \rangle &=& \frac{N}{4 \pi}
%(\d_{uv} \mp \e_{uv})
%\left[-h^{v \pm \a} \pm 2 Z^{v (+) \pm \a}  \right]_{0}, \nn
%\eea
%with $N = n_1 n_5$. The notation  $[A]_k$ indicates the coefficient
%of the $z^k$ terms  of A, where $z$ is the Fefferman-Graham coordinate,
%and $a_{Iij}=1/(2 \pi^2)\int Y^I Y^i Y^j$ is the triple overlap of harmonics.
%The overall normalizations (which lead to the specific overall numerical
%constants) are fixed by standard choices (as explained in
%\cite{KST}).
%
The derivation of these (and other) holographic 1-point functions
is discussed in detail in \cite{KST}. 
%Here we only make a few
%remarks. The terms linear in the fields are what one would anticipate based
%on the linearized analysis in \cite{Deger:1998nm}, except that in that
%paper the spectrum was computed in the de Donder gauge but the solution
%in (\ref{ez1}) is not in this gauge. This leads to the dependence
%on the $\r$ field. The non-linear terms originate both from non-linear
%terms in the KK map from six dimensions to three dimension and from
%non-linear terms in the relations between 1-point functions and
%radial canonical momenta due to extremal couplings \cite{Skenderis:2006uy}.
%
%Thus 
To compute the explicit value of the vevs we need to
expand asymptotically the solution in (\ref{ez1}).
The (near-horizon) harmonic functions appearing in the solution
can be expanded as
\bea
f_5 &=&
\frac{Q_5}{r^2} \sum_{k,I} \frac{f^{5}_{kI} Y^I_k(\q_3)}{r^k}; \quad
f_1 =
\frac{Q_1}{r^2} \sum_{k,I} \frac{f^{1}_{kI} Y^I_k(\q_3)}{r^k};
\nn \\
A_i &=& \frac{Q_5}{r^2}
\sum_{k \ge 1,I} \frac{(A_{kI})_i Y^I_k(\q_3)}{r^k}, \la{p0}
\eea
where $(r,\theta_3)$ are polar coordinates in the overall transverse
$R^4$ and we explicitly indicate the degree of the harmonics $k$.
The coefficient $f^{5}_{kI}, f^{1}_{kI}, (A_{kI})_i$ for the harmonic
functions in (\ref{mathu1}) will be given below; for now we proceed
in general. Inserting these expansions into the solution and extracting
the relevant coefficients leads after some manipulations to
\bea \label{vevs}
\langle j^{\pm\a} \rangle  &=& \pm
\frac{N}{2 \pi} (A_{1j})_i e^\pm_{\a ij} d y^\mp,\quad
\langle O_{S^1_i} \rangle
= -\frac{N \sqrt{2}}{\pi} f^{5}_{1i}; \nn \la{vv2} \\
\langle O_{\Sigma^2_I} \rangle &=& \frac{N \sqrt{2}}{4 \pi}
( -  (f^1_{2I} +f^5_{2I}) + (A_{1j})_i (A_{1l})_k e^I_{ijkl}),
\eea
where $y^{\pm} {=} y {\pm} t$, $I \equiv i$ and $I_v \equiv \a$ 
for degree 1 harmonics,
$e^I_{ijkl}{=}8 f^I_{\a \b} e^+_{\a ij} e^-_{\b kl}$, 
$f^I_{\a \b}{=}1/(2 \pi^2)\int Y^I (Y^{\a -})^a (Y^{\b +})_a$ and
$e^\pm_{\a ij}{=}\sqrt{Q_5/Q_1}/(2 \pi^2)
 \int (Y^{\a \mp})^b Y^j \partial_b Y^i$.
Notice that the vevs of the operators of dimension 1 depend linearly
on the degree 1 coefficients in (\ref{p0}) and the vevs of
the dimension 2 operators depend linearly on the degree 2 coefficients and
quadratically on the degree 1 coefficients. This is the structure
of the vevs of all chiral primary operators, i.e.
the vev depends linearly on the degree $k$
coefficient, where $k$ is the dimension of the operator,
and non-linearly on lower degree coefficients but such
that the sum of their degrees is $k$. One can easily obtain
the linear terms but computation of the coefficients of the non-linear terms
is very tedious in practice, although it is in principle possible,
and has only been done up to dimension 2 operators \cite{KST}.
This concludes our general discussion of the vevs associated with the solution
(\ref{ez1}).

We now consider the solutions specified by a closed curve
$F^i$.
In this case the coefficients (relevant for us) in (\ref{p0}) are given
by
\bea \la{p00}
f^5_{kI} &=& \frac{1}{ (k+1) L}
\int_{0}^{L} dv C^{I}_{i_1 \cdots i_k} F^{i_1} \cdots F^{i_k}; \\
f^1_{kI} &=&  \frac{Q_5}{ Q_1 (k+1) L}
\int_{0}^{L} dv | \dot{F} |^2
C^{I}_{i_1 \cdots i_k} F^{i_1} \cdots F^{i_k}; \nn \\
(A_{1j})_i &=& \frac{1}{2 L}
\int_{0}^{L} dv \dot{F}_{i} F_j. \nn
\eea
where $C^I_{i_1 \cdots i_k}$ are the orthogonal symmetric
traceless rank $k$ tensors on $R^4$.

The operators $O_{S^1_i}$ and all but one of the operators
$O_{\Sigma^2_I}$ are charged under the Cartan of $SO(4)$.
It follows that for the solutions to correspond to
a R ground state the vevs of these operators must vanish.
Before discussing the general case, let us consider the case of
an ellipsoidal planar curve in the 1-2 plane,
\be \label{ell}
F^1 = \frac{\mu \a}{n} \cos \frac{2 \pi n v}{L}, \quad
F^2 = \frac{\mu \b}{n} \sin \frac{2 \pi n v}{L},
\ee
where $\mu=\sqrt{Q_1 Q_5}/R$ and $(\a,\b)$ are parameters
satisfying
%For this curve (\ref{one-brane}) requires that
$\a^2 + \b^2=2$. Evaluating the vevs of the charged scalar operators
we find the following non-zero vevs,
\be \label{ch_vev}
\langle O^{(\pm 2,0)}_{\Sigma^2} \rangle = -\frac{\sqrt{3} N}{8 \sqrt{2} \pi}
\frac{\m^2}{n^2} (\a^2 -\b^2),
\ee
where we have explicitly indicated the $U(1) \times U(1)$ charges
$(j^R_3 + \bar{j}^R_3, j^R_3 - \bar{j}^R_3)$.
The vev of the R-symmetry current
is $\langle j^{\pm 3} \rangle  = (N/4 \pi) (\m/n)\a \b d y^{\mp}$.

We thus find that only for circles (with $\a^2 = \b^2=1)$)
do the vevs of charged operators vanish and therefore only these geometries
can correspond to R ground states. One might have anticipated this
result: a planar ellipse breaks the $SO(4)$ R-symmetry to $SO(2)$
(times certain discrete transformations) and this is reflected in the vev of
$O^{(\pm 2,0)}_{\Sigma^2}$. The vanishing of the vevs
of $O_{S^1_i}$ and of the other 6 charged operators
$O_{\Sigma^2}$ is due to the preserved $SO(2)$ and discrete symmetries.
It is then clear that the only geometries that have vanishing vevs
for all charged operators are the ones that have an $SO(2) \times SO(2)$
symmetry. The only such solutions are the circular ones.
(There may be non-circular curves with
$f_{1i}^5=f_{2I}^1=f_{1I}^5=0$ so that all vevs in (\ref{vevs}) are
zero but in this case one would find non-zero vevs of higher dimension
charged operators characterizing the symmetry breaking).

Thus we arrive at the conclusion that of all fuzzball solutions
only the circular ones can correspond to R ground states.
This seems to be at odds with the fact that all solutions
(\ref{ez1})-(\ref{mathu1}) are related by dualities \cite{Lunin:2001jy}
to solutions that are believed to correspond to strings carrying momentum
and thus should be related to bound states of D1 and D5 branes.
This argument, however, only implies that each geometry should be
related to some linear superposition of R ground states. Indeed,
the vevs of charged operators in a state that is not an eigenstate
of the R-charge are allowed to be non-zero. 

We now propose a map between fuzzball geometries and
superpositions of R ground states, which passes all kinematical and all
accessible dynamical tests. The motivation for this map and
various tests are discussed in detail in \cite{KST}.
Given the curve $F^i(v)$ which defines the
fuzzball geometry, we construct a corresponding coherent state in the
dual FP system and then find which Fock states in this coherent state
satisfy the constraint $N_L = N$, where $N_L$ is the left moving 
excitation number. Mapping these states to the D1-D5 system,
corresponding fundamental string oscillators to R operators, yields
the superposition of ground states which is proposed to be dual to the
D1-D5 geometry. 

In particular, for the case of the ellipse in (\ref{ell}), the conjectured
superposition is
$|ellipse)  =  \sum_{k=0}^{N/n} 2^{-N/n} \sqrt{C(N/n,k)}%a_k 
(\a+\b)^{\frac{N}{n}-k}
(\a-\b)^k {\cal {O}}^R_k,$ 
where $C(N/n,k)$  is the bimomial 
coefficient and 
$ {\cal O}^R_k$ are particular R ground states with R charge
$j^R_3 = \bar{j}^R_3 = (N/2n - k)$. This proposal passes all
kinematical checks and yields the correct answer for the vev of the 
R-symmetry current quoted above. 
The field theory computation of the vev in (\ref{ch_vev}) as well
as the vevs of other neutral and charged operators 
would require a knowledge of certain multiparticle three point functions at
strong coupling, which are not known, but approximating these
leads to vevs which agree remarkably well with
those extracted from supergravity. In particular, one recovers 
(\ref{ch_vev}) up to an overall numerical factor.

This then leaves the question: ``what are the 
geometries that are dual to R ground states''?
As has already been discussed,
a geometry which is dual to a R ground state must preserve the
$SO(2) \times SO(2)$ symmetry. This immediately implies that the asymptotics
must be of the following form:
\bea \la{av-a}
&&f_5 = \frac{Q_5}{r^2} \sum_{k=2l} \frac{f^5_{k0}}{r^k} Y^{0}_{k}; \quad
f_1 = \frac{Q_1}{r^2} \sum_{k=2l} \frac{f^1_{k0}}{r^k} Y^{0}_{k}, \\
&&A_a = \sum_{k} \frac{Q_5}{r^{k+1}} (A_{k0+} Y^{0+}_{ka} + A_{k0-}
Y^{0-}_{ka}); \nn
\eea
where the scalar spherical harmonics $Y^0_{2l}$ are
singlets under $SO(2) \times SO(2)$,
and $Y^{0 \pm}_{ka}$ are vector spherical harmonics of degree $k$
($k$ odd) whose Lie derivatives along the $SO(2)^2$ directions are
zero; note that these forms have only
components along the $(\phi,\psi)$ directions. 

Several examples of exact solutions which
have such asymptotics are discussed in \cite{KST}. 
One class of such solutions can be obtained
by superimposing harmonic functions, sourced on the $SO(2) \times SO(2)$
orbits of a given curve. 
%By construction such solutions will be symmetric under $SO(2) \times SO(2)$. 
Another class of solutions is
obtained using disconnected circular curves, although one would expect
that such solutions do not correspond to Higgs branch vacua. 
All of these solutions
pass the kinematical constraints to be identified as the dual of
R ground states (or chiral primaries after spectral flow).
To distinguish between them, one will thus need to use dynamical
information, namely the actual non-zero values of the vevs of neutral
operators.  This is a very subtle issue which will in general require
going beyond the leading supergravity approximation, as we discuss
in \cite{KST}.

In conclusion, our results support the overall fuzzball picture: the solutions
can be in correspondence with the black hole microstates in a 
way that is compatible with the AdS/CFT correspondence. The detailed
correspondence however is more complicated than anticipated.
Furthermore, even in the simplest 2-charge system one would need
to go beyond the leading supergravity approximation to test any 
proposed correspondence.


\begin{thebibliography}{99}

\bibitem{Lunin:2001jy}
  O.~Lunin and S.~D.~Mathur,
  Nucl.\ Phys.\ B {\bf 623}, 342 (2002),
  hep-th/0109154.

\bibitem{Mathur:2005zp}
  S.~D.~Mathur,
hep-th/0502050.

\bibitem{Lunin:2002bj}
  O.~Lunin, S.~D.~Mathur and A.~Saxena,
  Nucl.\ Phys.\ B {\bf 655} (2003) 185,
hep-th/0211292.

%\bibitem{Taylor:2005db}
%  M.~Taylor,
%  JHEP {\bf 0603} (2006) 009,
%  hep-th/0507223.

\bibitem{Lunin:2002iz}
  O.~Lunin, J.~Maldacena and L.~Maoz,
hep-th/0212210;
M.~Taylor,
  JHEP {\bf 0603} (2006) 009,
  hep-th/0507223.

\bibitem{Dabholkar:1995nc} C.~G.~Callan, J.~M.~Maldacena and A.~W.~Peet,
  Nucl.\ Phys.\ B {\bf 475}, 645 (1996)
  [arXiv:hep-th/9510134];
  A.~Dabholkar, J.~P.~Gauntlett, J.~A.~Harvey and D.~Waldram,
  Nucl.\ Phys.\ B {\bf 474}, 85 (1996), hep-th/9511053.

\bibitem{KST} I.~Kanitscheider, K.~Skenderis and M.~Taylor, hep-th/0611171.



\bibitem{Gubser:1998bc}
  S.~S.~Gubser, I.~R.~Klebanov and A.~M.~Polyakov,
  Phys.\ Lett.\ B {\bf 428}, 105 (1998),
  hep-th/9802109;
  E.~Witten,
  Adv.\ Theor.\ Math.\ Phys.\  {\bf 2}, 253 (1998),
  hep-th/9802150.

\bibitem{deHaro:2000xn}
  S.~de Haro, S.~N.~Solodukhin and K.~Skenderis,
  Commun.\ Math.\ Phys.\  {\bf 217}, 595 (2001),
  hep-th/0002230,
 

\bibitem{Skenderis:2002wp}
  K.~Skenderis,
  Class.\ Quant.\ Grav.\  {\bf 19}, 5849 (2002),
  hep-th/0209067.

\bibitem{Skenderis:2006uy}
  K.~Skenderis and M.~Taylor,
  JHEP {\bf 0605}, 057 (2006),
  hep-th/0603016.

\bibitem{Deger:1998nm}
  S.~Deger, A.~Kaya, E.~Sezgin and P.~Sundell,
  Nucl.\ Phys.\ B {\bf 536}, 110 (1998),
  hep-th/9804166.

\bibitem{deBoer:1998ip}
  J.~de Boer,
  Nucl.\ Phys.\ B {\bf 548}, 139 (1999),
  hep-th/9806104.


\end{thebibliography}
\end{document}